\documentclass{article}
\usepackage{newtxtext,newtxmath}

\usepackage{graphicx}
\usepackage{hyperref}
\usepackage[letterpaper,margin=1in]{geometry}

\usepackage{comment}

\renewenvironment{abstract}
	{\quotation}
	{\endquotation}

\date{}

\makeatletter
\renewcommand{\fnum@figure}{\textbf{Figure \thefigure}}
\renewcommand{\fnum@table}{\textbf{Table \thetable}}
\makeatother

\usepackage{scicite}

\usepackage{url}
\usepackage{authblk}

\graphicspath{{}}

\newcommand*{\GdMn}{GdMn$_2$O$_5$}
\newcommand{\bm}[1]{{\mathbf #1}}

\definecolor{ForestGreen}{RGB}{34,139,34}
\definecolor{Orchid}{RGB}{218,112,214}

\def\scititle{
    Highly controllable switching pathways in multiferroic GdMn$_2$O$_5$
}
\title{\bfseries \boldmath \scititle}

\author[1]{M. Ryzhkov\thanks{These authors contributed equally}}
\author[6]{A. Granero$^{\displaystyle*}$}
\author[1]{J. Wettstein}
\author[1]{Anna Pimenov}
\author[2]{X.~Wang}
\author[5]{L. Ponet}
\author[3]{S.-W. Cheong}
\author[4]{M. Mostovoy}
\author[1]{Andrei Pimenov
}
\author[7]{S. Artyukhin\thanks{Corresponding author: sergey.artyukhin@gmail.com}}

\affil[1]{Institute of Solid State Physics, Vienna University of Technology, 1040 Vienna, Austria.}
\affil[2]{University of Science and Technology, Beijing, China.}
\affil[3]{Rutgers Center for Emergent Materials and Department of Physics and Astronomy, Rutgers University, New Jersey 08854, USA.}
\affil[4]{Zernike Institute for Advanced Materials, University of Groningen, the Netherlands.}
\affil[5]{Theory and Simulation of Materials (THEOS) and National Centre for Computational Design and Discovery of Novel Materials (MARVEL), École Polytechnique Fédérale de Lausanne, Lausanne 1015, Switzerland.}
\affil[6]{Quantum Materials Theory, Istituto Italiano di Tecnologia, 16163 Genova, Italy.}
\affil[7]{Quantum Materials Theory, 16162 Genova, Italy.}

\begin{document} 

\maketitle

\begin{abstract} \bfseries \boldmath

Controlling magnetic moments using electric fields remains a central challenge in spintronics. Multiferroics, where magnetic and electric orders coexist, may be a natural platform for such control, but progress has been limited because interactions between these orders are typically too weak to overcome the energy barriers between magnetic states. A recently demonstrated topologically protected switching circumvents this limitation by exploiting reduced barriers at a phase transition. Nevertheless, the conditions enabling this phenomenon remain elusive and electric field control is poorly understood.
Here, we experimentally map the energy landscape by tracking transitions in GdMn$_2$O$_5$ under combined electric and magnetic fields. The experiments reveal that the switching pathways can be controlled by the electric field. A minimal theoretical model captures the observed behavior, identifies the parameter space where switching paths are sensitive to small perturbations and reveals design principles for implementing topological switching in other materials.


\end{abstract}

\noindent
\subsection*{Introduction} In the race to meet the ever-growing demands on data storage, both in capacity and performance, miniaturization has always been a sure-fire way to progress technology in these two areas.
However, as device dimensions have decreased, certain undesirable effects have been exacerbated. These then led to increased error rates and ultimately limited the effectiveness of further miniaturization.
This causes researchers to look for radically different data storage and retrieval mechanisms in an effort to engineer the next giant leap.
Recent work on multiferroic materials has paved a promising path \cite{Spaldin2019,Khomskii2009,Fiebig2005,Fiebig2016,Cheong2007}.
Magnetoelectric multiferroics harbor two coupled orders, magnetism and ferroelectricity, implying the possibility to control one order by perturbing the other.
Making it possible to switch the magnetic order using an electric field could break the ground for higher density and low-consumption data storage devices.
In general, this is challenging due to the different symmetries of magnetic and ferroelectric orders, but simultaneous breaking of inversion and time-reversal symmetries in multiferroics may allow magnetoelectric effects \cite{Bibes2008,Kleemann2013,Heron14,Matsukura2015,Manipatruni2018}.

Even then, the electric control of magnetism was achieved only in a handful of single-phase multiferroics, e.g. BiFeO$_3$, orthoferrites, and hexaferrites \cite{Heron14, Tokunaga2012, Zhai2017giant, he2010robust, erickson2024imaging}. Low magnetically induced electric polarization does not allow one to control spins by an applied electric field. Even in GdMn$_2$O$_5$ boasting one of the highest ferroelectric polarizations, $P = 3.5\times10^3$ $\mu{\rm C}/{\rm m}^{2}$, measured in type II multiferroics, the magnetoelectric energy for $E = 5\times10^4$ ${\rm V}/{\rm cm}$ is about an order of magnitude lower than the magnetic anisotropy energy and 2-3 orders of magnitude lower than the exchange energy. Here, we demonstrate a rich variety of switching pathways made possible by an external $E$-field in GdMn$_2$O$_5$ when the $H$-field is ramped across the spin reorientation transition. We rationalize the behavior by a minimal model, which clarifies the design principles for topological switching and the requirements for electric spin manipulation at magnetic transitions in multiferroics.


%
\begin{figure*}
    \centering
    \includegraphics[width=1\linewidth]{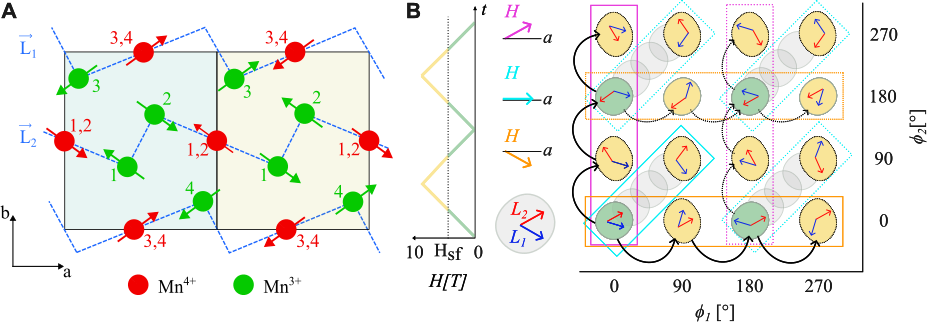}
    \caption{\label{fig:fig1}{\bf $E$-field modification of topological switching} (\textbf{A}) Magnetic unit cell of \GdMn. Mn$^{4+}$ ions (red spheres) and Mn$^{3+}$ ions (green spheres) form zigzag antiferromagnetic chains (blue dashed lines) running along the $a$-axis. $\mathbf{L}_1$ and $\mathbf{L}_2$ are the N\'eel vectors of the two inequivalent chains. 
    Arrows indicate a ground state spin configuration, which doubles the crystallographic unit cell along $a$, with opposite spins in the neighbouring cells. 
    (\textbf{B}) Magnetic field ramp protocol (left) and schematic energy landscape in the space spanned by the angles $\phi_{1,2}$  between the $a$-axis and $\mathbf{L}_{1,2}$ vectors. Yellow (green) puddles represent local energy minima at $H$ field below (above) the spin reorientation transition. $\mathbf{L}_1$ and $\mathbf{L}_2$ are indicated by blue and red arrows in each puddle. Purple boxes indicate the possible transition paths when $H$ field is applied at a positive angle to the $a$-axis  ($\Phi_H > 0$), driving the switching along vertical valleys, while orange boxes show the path followed for $\Phi_H < 0$, where horizontal valleys are formed. Black arrows illustrate the single switching events occurring when the $H$ field is ramped across the spin reorientation field $H_{sp}$. Switching at zero $E$-field goes along these vertical (horizontal) valleys when $\Phi_H$ is inside the magic angle region, $\Phi_H = + \Phi_H^*$ ( $\Phi_H = - \Phi_H^*$), while outside an external $E$-field is necessary to drive these transitions. Blue boxes highlight the state deformation in which the former local minimum shifts into the following one where $\mathbf{L}_1$ and $\mathbf{L}_2$ are symmetrical with respect to the $a$-axis when the $H$ field is applied along the $a$-axis. Here, again, a degree of control of the transition path is possible under a relatively small $E$-field (cf. Fig.~\ref{fig:fig3}{\bf A,B,E,F}).}
\end{figure*}

\GdMn\ is a magnetically frustrated multiferroic material, in which ferroelectricity is induced by inversion-breaking AFM order. The antiferromagnetic ordering in the two zigzag spin chains is described by two N\'eel vectors, $\bm{L}_1$ and $\bm{L}_2$, one of which is even and another is odd under inversion, which forbids the linear coupling between them, but allows ferroelectric polarization $P_b\propto \bm{L_1}\cdot\bm{L_2}$. This frustration implies that the two N\'eel vectors may be switched independently.

The crystal lattice of the multiferroic \GdMn\ is centrosymmetric and belongs to the space group P$bam$ (No. 55) ~\cite{Alonso97a}. 
Octahedrally coordinated Mn$^{4+}$ ions and pentahedrally coordinated Mn$^{3+}$ ions form zigzag chains along the $a$-axis (blue lines in Fig.\,\ref{fig:fig1}{\bf A}) with antiferromagnetic (AFM) exchange interactions between neighboring spins along the chains. Mn ions from neighboring chains form AFM pentagons with relatively weak and geometrically frustrated interchain exchange interactions~\cite{Chapon04,Chapon06,Kim11}.
At $T_{\rm N1}=40$~K GdMn$_2$O$_5$ orders magnetically, adopting an incommensurate state with the propagation vector $\mathbf{q} = (0.49, 0, 0.18)$.
Below $T_{N2}=33$~K it locks into a commensurate state with the unit cell doubling along the $a$-axis,
showing one of the highest magnetically induced electric polarizations of $P_b \sim 3 600$\,$\mu$C/m$^2$, and a large variation (up to 5000\,$\mu$C/m$^2$) in an applied magnetic field \cite{Lee13}. This field-induced polarization variation
is robust and changes very little when multiple field sweeps are applied. 
Electric polarization originates from the symmetric exchange striction in Mn-Mn and Mn-Gd bonds and has a large electronic component~\cite{Lee13,Giovannetti08}.

Recently, a topologically protected switching was discovered in \GdMn\ \cite{Ponet2022}. 
When the magnetic field is directed at approximately $\Phi_H \approx \pm 10^\circ$ to the $a$ axis, and is ramped up and down twice, the system traverses four distinct polarization states, labeled 1,2,3, and 4 (Fig.~\ref{fig:fig1}). This sequence resembles a mechanical crankshaft, as in these experiments the magnetic field ramp causes half of Mn and Gd spins to rotate clockwise or counterclockwise, depending on the $H$-field orientation.
The trajectory of the point $(\phi_1(t),\phi_2(t))$ describing the evolution of order parameter directions in this regime is incontractible on $(\phi_1,\phi_2)$ torus (Fig.~\ref{fig:fig1}{\bf B}, Fig.~\ref{fig:fig3}{\bf A}) \cite{Ponet2022}. The winding number, defined as the change of the order parameter rotation angle divided by $2\pi$, is an integer topological invariant, equal $\pm 1$ in this regime. Being an integer, the topological invariant guarantees exactly $2\pi$ spin rotation in some parameter region, conveying topological protection to the switching process. Such topologically protected four-state loop is only observed in a narrow sector of magnetic field orientations, $\Phi_H \approx \pm\Phi_H^*= 10^\circ$. At a positive $\Phi_H$, $L_2$ rotates in increments of approximately $90^\circ$ through each field ramp, as the system visits the states in a vertical magenta box in Fig.~\ref{fig:fig1}{\bf B}, while at a negative $\Phi_H$, $L_1$ rotates and the states in a horizontal orange box are visited. 
$|\Phi_H| < \Phi_H^*$, the electric polarization is reversed by a magnetic field. For $\Phi_H>\Phi_H^*$, the sign of the electric polarization does not change during the whole magnetic field sweep \cite{Ponet2022}.

Here, we show that magnetoelectric switching between the four states can be efficiently controlled with an external {\em electric} field (Fig.\,\ref{fig:1234}), allowing us to reach any desired magnetoelectric state. Importantly, we can electrically reverse the switching direction for a wide range of magnetic field orientations, bypassing the requirement for the magnetic field angle to be within the magic interval. An additional electric field contribution to the energy balance of the system enables a degree of control of the electric polarization and magnetic state during a magnetic field sweep. The schematic energy landscape with switching paths shown in Fig.~\ref{fig:phaseDiag}{\bf C} demonstrates that all four distinct polarization states can be reached outside the ``magic'' angle region~($|\Phi| < \Phi_H^*$) in a sequence of steps.

\subsection*{Experimental results and discussion} We recall that in the two ``magic" angle regimes near $\Phi_H \approx \pm 10^\circ$ the system runs through a predefined sequence of magnetoelectric states: 1-2-3-4-1 ($\Phi_H = -10^\circ$) or 1-4-3-2-1 ($\Phi_H = +10^\circ$), respectively. At $\Phi_H = 0$, the switching goes between states 1 and 4 and persists up to 25~K (Fig.~S3). We have experimentally proved that these two sequences are robust with respect to the poling voltage, i.e. the application of a moderate external electric field $E \sim 500$~V/mm cannot break the order of the states. However, the follow-up experiments demonstrated that for external magnetic fields close to the $a$-axis the switching by electric field becomes possible. Indeed, two low-field states have opposite $P_b$, as do two high-field states, thus being able to control the final state with $E_b$ may be naively expected at least in some range of parameters. 

\begin{figure*}[t!]
    \centering    \includegraphics[width=1\textwidth]{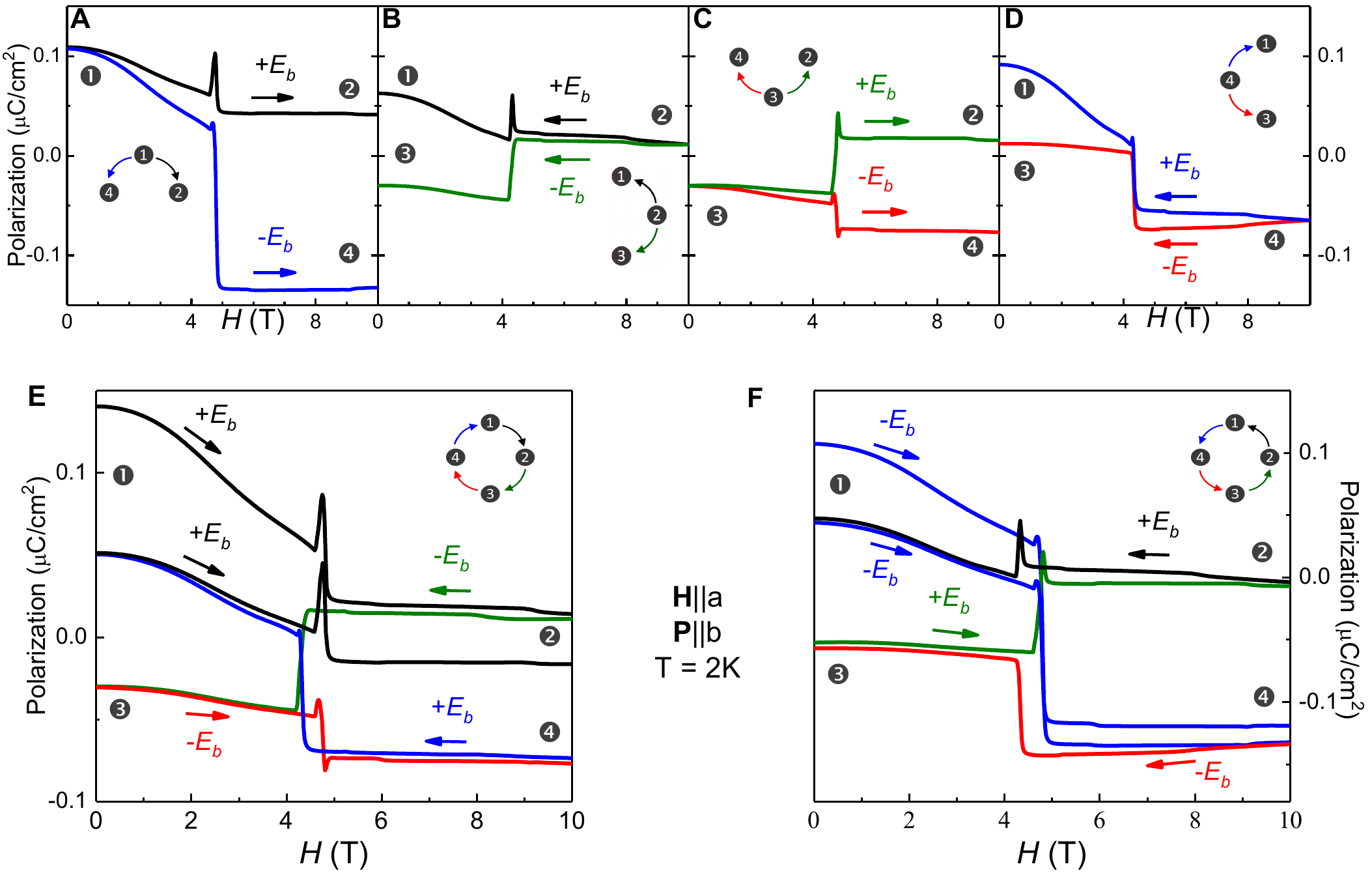}
    \caption{\label{fig:1234}{\bf E-field switching of magnetoelectric states in GdMn$_2$O$_5$.} (\textbf{A-D}) E-field controlled switching between single unit states of the system. (\textbf{A}) Switching from state 1 to states 2 and 4 at positive and negative applied voltage, respectively. (\textbf{B}) Switching from state 2 to states 1 and 3. (\textbf{C}) Switching from state 3 to states 2 and 4. (\textbf{D}) Switching from state 4 to states 3 and 1. (\textbf{E, F}) Combined four-state switching sequence with winding controlled through external electric fields: (\textbf{E}): clockwise cycle 1-2-3-4-1. (\textbf{F}): anti-clockwise cycle 1-4-3-2-1. Magnetic field is applied parallel to the a-axis (nominal deviation $\sim 1^\circ$).}
\end{figure*}

In fact, this new scheme can be realized. The summary of the characteristic experiments is given in Fig.~\ref{fig:1234}. First, we recall  that states 1,3 correspond to zero magnetic field and states 2,4 are high-field states. As the switching procedure includes the ramping of the magnetic field, only the transitions between two zero-field and two high-field states are possible. The corresponding set of possible transitions, realized experimentally, is shown in Fig.~\ref{fig:1234}{\bf A-D}. Starting from state 1 (panel {\bf A}) the system can be moved either to state 2, applying positive voltage (black curve), or to state 4, applying negative voltage (blue curve). Similarly, from state 2, the zero-field states 1,3 can be reached applying positive or negative voltage, respectively (panel {\bf B}). Finally, panels {\bf C,D} demonstrate that two other transitions, i.e. from 3 to 2,4 and from 4 to 1,3, can be obtained as well. 

Because all possible single steps are available with $E$ field control, we can now implement the 4-cycle switching both clockwise 1-2-3-4-1 (Fig.\,\ref{fig:1234}{\bf E}) and counter-clockwise 1-4-3-2-1 (Fig.\,\ref{fig:1234}{\bf F}). These two cycles correspond to the original topologically protected switchings at ``magic" tilting angles~\cite{Ponet2022} of the external magnetic field. The general procedure in both switching cycles is the following: Two magnetoelectric states 1,2 with high polarization are always reached with a positive applied voltage, whereas states 3,4 with low (or negative) polarization are reached by application of a negative voltage. This agrees with the mechanism described above, for which the $E$ field is the crucial element for the system to decide between different states.

The four-state sequences in Fig.~\ref{fig:1234}{\bf E,F} clearly show that the polarization does not return to the initial value after a full cycle. This is especially true for experiments that begin with the polarization state 1 as obtained by an initial cooling run in the poling field. However, in subsequent cycles, the polarization values of states 1-4 stabilize. This training effect is likely related to the formation of a multi-domain polarization state. It also manifests in states 1 and 3 (and 2 and 4) not having opposite ferroelectric polarizations.

The electric field applied for the measurements shown in Fig.~\ref{fig:1234} was $380\,$V/mm. The induced electric polarization due to the dielectric permittivity $P_{b,\mathrm{ind}} = \varepsilon_0 \chi E$ was subtracted from the graphs shown in figure \ref{fig:1234} for clarity. Here, $\varepsilon_0$ and $\chi = 17 \pm 2$ are the vacuum permittivity and the
dielectric constant of \GdMn,\cite{bukhari_prb_2016}, respectively.


\subsection*{Modelling} Prior modelling of topological switching relied on a microscopic model which included individual Mn and Gd spins. The assumption of strong AFM exchange in zigzag Mn chains traversing the unit cell along the crystallographic $a$ direction allowed to replace individual Mn spins by two rigid AFM orders ${\bf L}_1, {\bf L}_2$ in the two chains \cite{Ponet2022}. Here, we find that an even simpler model including only these two AFM chains can capture topological switching.
%


\begin{figure}[b!]
    \centering
    \includegraphics[width=\linewidth]{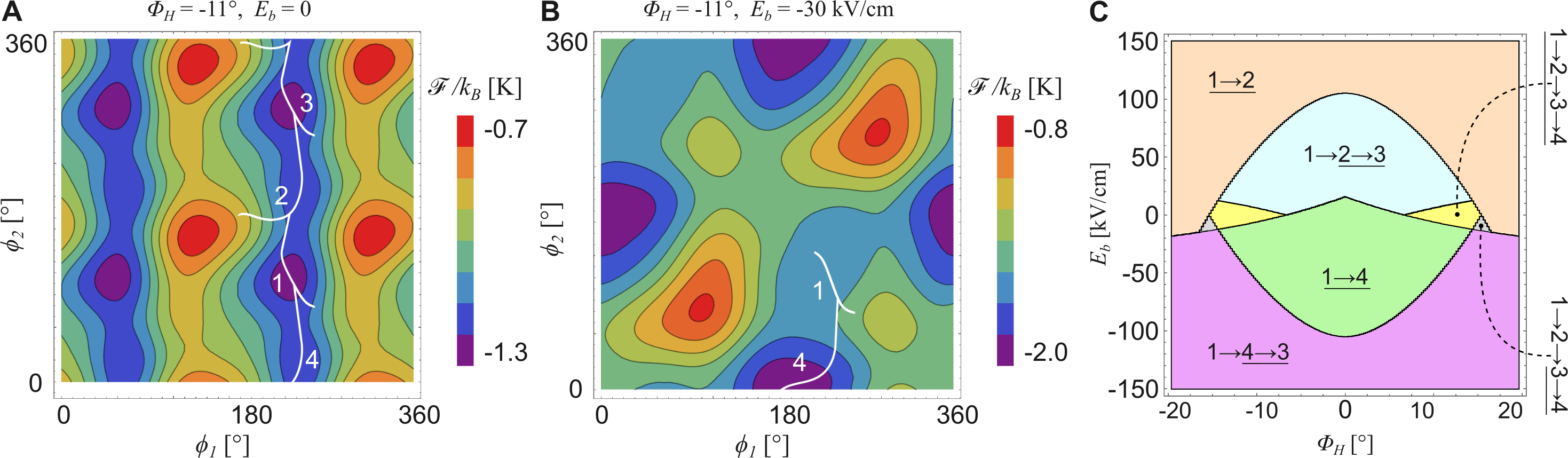}
    \caption{\label{fig:phaseDiag} {\bf Energy landscape probed by electric field and resulting in a rich switching diagram} (\textbf{A}) The trajectory in the space of order parameter orientation $(\phi_1(t),\phi_2(t))$, corresponding to topological switching at $\Phi_H=-11^\circ$ and $E_b=0$, overlaid on the potential energy surface for $H=4.5$~T. The low-energy valleys run vertically. The numbers mark the four states, connected by $L_2$ rotation by $90^\circ$. (\textbf{B}) An applied $E_b$ field leads to diagonal valleys and breaks the topological switching sequence. (\textbf{C}) Simulated switching regimes starting from state 1 when the magnetic field at $\Phi_H$ to $a$-axis is swept up and down across the spin reorientation transition in the presence of an electric field $E_b$. The state sequences are indicated, with periodically repeating states underlined (e.g. $1\to \underline{ 2\to 3}$ implies the sequence 1232323\dots). }
    
\end{figure}

\begin{figure*}[t]
    \centering
    \includegraphics[width=\linewidth]{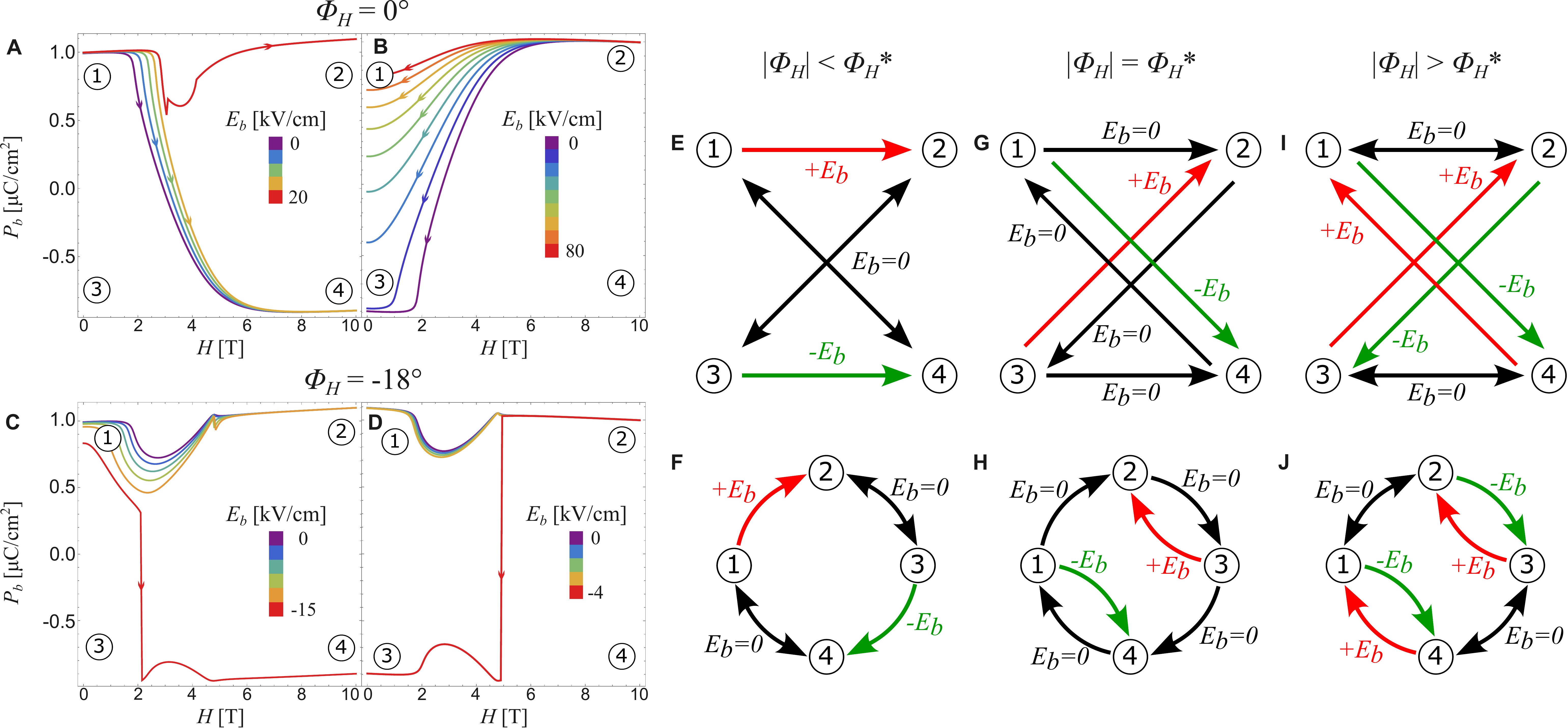}
    \caption{\label{fig:fig3}{\bf Simulated switching between magnetoelectric states under $H$ sweeps and a low constant $E$-field.} Color encodes the magnitude of $E_b$. (\textbf{A,B}) Switching driven by magnetic field sweeps at $\Phi_H = 0^{\circ}$. (\textbf{A}) At low $E$-field switching follows $1\to 4$, at high --- $1\to 2$. (\textbf{B}) Switching starting from state 2. Continuous deformation of $P(H)$ curves with increasing $E$-field indicates that the final magnetic state resembles state 3, strongly polarized by $E$-field. (\textbf{C, D}) $E$-field controlled switching starting from states 1 and 2, respectively, when the $H$-field angle $\Phi_H = -18^{\circ}$ is above the magic angle interval. 
    (\textbf{E-J}) Switching paths below (\textbf{E,F}), within (\textbf{G,H}) and above (\textbf{I,J}) the magic angle interval $7.7^{\circ} \leq |\Phi_H^*| \leq 17.7^{\circ}$. Switching paths starting in states 3 and 4 are obtained by a mirror $P\to -P, E\to -E, \bm L_1\to -\bm L_1$.}
\end{figure*}

Building invariants under the transformations of P$bam$ space group (\#55 in International Tables~\cite{brock2016international}), and retaining the important terms, we get the free energy per crystallographic unit cell, containing 4 formula units, of the form (details in SI):
\begin{equation}
\label{eq:F}
    \mathcal{F}= \gamma(\bm{L}_1\cdot\bm{L}_2)^2-\alpha E_b(\bm{L}_1\cdot\bm{L}_2) 
    +\sum_{i=1,2}\left[-K_2(\bm{L}_i\cdot \bm{n}_i)^2-K_4(\bm{L}_i\cdot \bm{n}_i')^4+\frac{\chi}{2}(\bm{L}_i\cdot\bm{H})^2\right]
,
\end{equation}
where the first term favors $\bm{L}_1\perp\bm{L}_2$ state that allows energy gain on interchain exchange interactions via spin canting~\cite{Sushkov08}. The second term represents the interaction of magnetically induced ferroelectric polarization $P_b=\alpha \bm{L}_1\cdot\bm{L}_2$ with the external electric field. Hence, rotations of AFM vectors are large-amplitude collective modes strongly coupled to the $E$ field. Since the term is proportional to $\cos(\phi_1-\phi_2)$, at high $E_b$ it drives the reconstruction of the free energy landscape, replacing vertical valleys, responsible for the topological switching (Fig.~\ref{fig:phaseDiag}{\bf A}), by valleys along the diagonal $\phi_1=\phi_2$ (Fig.~\ref{fig:phaseDiag}{\bf B}). The terms with $K_{2,4}$ represent the single-ion anisotropy for the two chains, with the easy axes $\bm{n}_i$ approximately along the $a$ axis. The Zeeman energy, gained on the spin canting in the magnetic field when the $\bm{L}_{1,2}\perp \bm{H}$, is represented by the terms with the coefficient $\chi$, whose scale is given by the inverse of the intrachain antiferromagnetic exchange. $\chi$ is further enhanced due to the interaction of the field with large Gd spins, closely aligned with Mn spins 
The term with $K_4$ is responsible for the width of the topological switching interval, and allows to model a wide $\Phi_H$ interval, observed experimentally. The values of the parameters are chosen to reproduce the experimental data (see SI). The topological region then extends in the range $7.7^{\circ} \le |\Phi_H| \le 17.7^{\circ}$. See Methods for the choice of model paramaters. 

This model reproduces the topological switching, observed in the ``magic'' angle region. For a magnetic field along the easy axis $\bf{n}_i$, the $K_2$ anisotropy and Zeeman terms compensate each other near the spin flop field $H\propto \sqrt{K_2J}$ \cite{oh14,Kim2015,Yokosuk16,Ono2024}. Thus, the leading anisotropy disappears, resulting in a rather flat valley in the potential energy surface along $\phi_i$ (Fig.~\ref{fig:phaseDiag}{\bf A}). That flatness remains even when the field is applied outside the ``magic'' angle interval.
The simulations reproduce the experimentally observed electric field control of switching at $\Phi_H = 0^{\circ}$ when starting from low-$H$ states, as seen in Fig.~\ref{fig:fig3}{\bf A} and in Supplementary Figure~\ref{fig:figS2}{\bf A,C}. 
We now turn to corroborate the role of the electric field in the switching behavior. 
Fig.~\ref{fig:phaseDiag}{\bf C} shows how the switching sequence is affected by the magnetic field angle and $E_b$. This switching diagram, showing how the switching pathways vary in the $(E,\Phi_H)$ plane, displays a variety of regimes. Its utility lies in the insights it provides to control the switching. In fact, by tuning the system to the boundaries between different regions, it is possible to alter the switching sequence by a small variation in $E$ or $\Phi_H$. Remarkably, towards the topological region, $E$-field control is substantially enhanced, i.e. a much lower $E$-field can alter the transition sequence. That is in line with the flattening of the energy surface in that regime.

However, when starting from high-$H$ configurations, the $E_b$ field polarizes the state that would be reached anyway in the $H$ sweep without $E_b$, as illustrated in Fig.~\ref{fig:fig3}{\bf B} (and in Supplementary Figure~\ref{fig:figS2}{\bf B,D}). The degree to which the external field polarizes the state is so extreme that the process could be interpreted as the $2\to 1$ transition at first glance, but the continuous deformation of the $P_b(H)$ curves with increasing $E_b$, seen in Fig. \ref{fig:fig3}{\bf B} suggests otherwise. Once $E_b$ is removed, the simulated polarization reverses to that achieved without $E_b$. In the experiments, the reversal is only partial and slow. We find $2\to 1$ switching to be possible only when the magnetic field is oriented near the magic angle interval and at much higher $E_b$.

The discussed switching sequences are summarized in Fig.~\ref{fig:fig3}{\bf E,F}. $E$-control allows us to reach all four states from any other state in a sequence of sweeps, as schematically described in Fig.~\ref{fig:fig3}{\bf E,F} (see Supplementary Figure~\ref{fig:figS2}{\bf A,C} for numerical simulations).

The model predicts different $E$-field controlled transitions when the magnetic field orientation is within the magic interval, where topological switching is observed at $E=0$. $E_b$ breaks topological sequence $\underline{1\to 2\to 3\to 4}$, inducing the $3\to 2$ transition when positive and $1\to 4$ when negative (Fig.~\ref{fig:fig3}{\bf G,H}). Here, the underlined states are periodically repeated under consequent field ramps.

When the field angle is above the magic interval, a moderate $E$-field changes the switching path, as indicated by the red curve in Fig.~\ref{fig:fig3}{\bf C,D}. This leads to a rich switching diagram in the $\Phi_H-E_b$ plane, Fig.~\ref{fig:phaseDiag}{\bf C}. The topological region (in yellow) is symmetric with respect to the reversal of $E_b$, and the $\Phi_H$ region shrinks with increasing $|E_b|$. In the gray area winding happens only once, followed by transitions between states 3 and 4 under subsequent magnetic field ramps.

The simulations explain how a moderate electric field enables the switching sequences presented in Figure~\ref{fig:1234} in a broad range of magnetic field orientations outside the ``magic" angle region, which was mandatory for the topological switching. Importantly, tuning towards the magnetic reorientation transition produces flat energy landscape which is the key for the electric control of magnetism.
It is worth to remark that the Landau-type model, developed here to explain both topological and $E$-controlled switching behavior in GdMn$_2$O$_5$, is entirely based on only two features: the P$bam$ space group symmetries and the two AFM manganese chains. Hence, it provides a basis for explaining analogous phenomena in other materials, such as those recently observed in the pioneering work to extend topological switching to other $R$Mn$_2$O$_5$ compounds\cite{Wang25}.

The model (1) 
is a good starting point for formulating the design principles for materials with topological switching. As seen in the simulations, the $K_4$ term broadens the magic angle interval, but is not essential for topological switching, as is the term with $E_b$. The other three terms constitute the minimal model. The essential ingredients are two antiferromagnetic order parameters that transform in different ways under the symmetry group, thus ruling out a bilinear term and allowing their independent rotations. Their easy axes $\bm{n}_{i}$ must be misaligned so that the magnetic field could cause a spin flop transition in one and not in the other, the latter guiding the next rotation with the same sense via the term with $\gamma$ (clockwise or counterclockwise). $\bm{L}_2$ in that term cannot just be replaced by a constant vector, leading to a tilted easy axis as that will lower e.g. the barriers $1-2$ and $3-4$. Instead, $\bm{L}_2$ tilts after the rotation of $\bm{L}_1$, now lowering the other pair of barriers, $2-3$ and $1-4$, and driving the topological switching further.
 

\subsection*{Conclusions}
Three factors make the electric control of magnetism in GdMn$_2$O$_5$ possible: (i) proximity to a spin-flop transition, (ii) magnetic frustration and (iii) relatively high magnetically induced electric polarization. Near the flop transition, the magnetic anisotropy becomes small, enabling $E$ field control of the switching pathway. Importantly, opposite parities of $\mathbf{L}_1$ and $\mathbf{L}_2$ exclude linear coupling between the two order parameters and frustrate interchain interactions. At the same time, the electric polarization strongly depends on the angle between $\mathbf{L}_1$ and $\mathbf{L}_2$, which makes it possible to toggle magnetic states with an applied electric field. 
Although the model reproduces the experimentally observed electric field control in the vicinity of the spin-flop transition, the electric field required to reverse the polarization direction in the experiments is much lower than that estimated in simulations. This discrepancy is to be attributed to the presence of depolarizing fields estimated as $4\pi P_0/\varepsilon \approx 230$~kV/cm, favoring a multidomain structure. Due to this depolarization field, near the spin-flop magnetic field, the energy barriers separating states with opposite polarizations are significantly reduced, leading to the broadening of the domain walls and the weakening of their pinning by lattice imperfections. A similar effect has been experimentally observed in GdMnO$_3$. This behavior results in broadened and weakly pinned domain walls that can be easily displaced by modest electric fields, favoring the growth of the domains with the polarization aligned along the applied field. The discrepancy between the measured switching field and that inside the ferroelectric could result from finite-width screening layer in the contacts \cite{dawber2003} and the charge accumulation at the interfaces during the switching procedure.

Inversion symmetry breaking is often a result of magnetic frustration that gives rise to unusual magnetic states and low-energy collective modes. It would be interesting to explore the electric control of magnetism in other multiferroics near spin-flop and spin-reorientation transitions. The results highlight tuning to the phase transition between two phases with different polarizations as a general material design principle allowing a robust electric switching in future technological applications. The vicinity to the boundaries of the switching diagram allows to alter the switching pathways by a small parameter variation. The minimal model for topological switching uncovers the essential ingredients and highlights that rare earth ions are not essential, and provides a blueprint for topological switching in other systems.

\section*{Acknowledgments}

\paragraph*{Funding:}
This work was supported by the Austrian Science Funds (I 5539-N 
and PAT7680623).
\paragraph*{Author contributions:} A.P. conceived the project. M.R., J.W., Anna P., have performed experiments under the supervision of Andrei P.; X.W. have grown the crystals under the supervision of S.-W. C.; A.G., M.M., L.P. and S.A. developed and simulated the model. S.A. supervised the modelling effort. All authors participated in the discussions and edited the text.
\paragraph*{Competing interests:}
The authors declare no competing interests.

\paragraph*{Data and materials availability:}
Experimental data presented in this paper are open access and can be found at \href{https://doi.org/10.5281/zenodo.18745858}. 


\clearpage 

%
\bibliographystyle{sciencemag}
\bibliography{gdmn2o5}
\clearpage
\subsection*{Materials and Methods}
\subsubsection*{Experimental}

Single crystals of \GdMn~ were grown by flux method \cite{Lee13} and  characterized by X-ray analysis and by electric, dielectric, and magnetic measurements. From these experiments the magnetoelectric phase diagram in external magnetic fields was obtained \cite{bukhari_prb_2016}.  Static electric polarization was measured on a small crystals of typical size $0.4\times0.4\times0.4$\,mm$^3$ and using silver paste for electric contacts.
The polarization was measured using an Keithley electrometer adapted to a Physical Property Measuring System, with magnetic fields of up to 14\,T and temperatures down to 1.8\,K. By changing the orientation of the sample in the cryostat, the direction of the magnetic field relative to the crystal axes was adjusted. The crystal orientation of the samples was determined using the Laue diffractometer. In order to reduce the mechanical torque due to the off-axis magnetic field, the maximum field values were limited to 10\,T.

\subsubsection*{Modelling}
Topological swithcing in GdMn$_2$O$_5$
has been modelled using a microscopic Hamiltonian, that included two AFM Mn chains and 8 Gd ions per unit cells. As a result, a large number of microscopic interactions were included. Here we derived a minimal Landau-type model, having the symmetry of P$bam$ paramagnetic space group (see Supplementary and Supplementary Table S1). We performed numerical simulations to explore the free-energy landscape and its evolution under external fields.
We parametrize the two antiferromagnetic order parameters by azimuthal angles $\phi_{1,2}$ as $\bm{L_{1,2}}=(\cos\phi_{1,2}, \sin\phi_{1,2}, 0)$, with easy-axis directions $\bm{n_{1,2}}=(\cos\phi_{1,2}, \sin\phi_{1,2}, 0)$, where $\phi_{1,2}=\pm 6^\circ$ respectively. The external magnetic field is written as $\bm{H}=h(\cos\Phi_H, \sin\Phi_H, 0)$. By fitting experimentally observed width of the topological switching region and the magnitude of the ferroelectric polarization, the following set of parameters was found: $\gamma = 0.133$~meV/cell, $\alpha = 0.01$~meV$\cdot$cm/kV$\cdot$cell, 
$K_2=0.314$~meV/cell, $K_4 = -0.837$~meV/cell, $\phi_2= 0.267$, $\phi_4=0.426$, $\chi=0.086$~meV/T$^2\cdot$cell, where ``cell'' refers to the crystallographic unit cell. These values are in agreement with the order of magnitude microscopic estimates, presented in the Supplementary.
The simulations explored the evolution of the potential energy surface as the magnetic-field strength was swept quasistatically across the experimental range $0T \leq h \leq 10T$ in $0.05 T$ steps. The calculations were done for different in-plane orientations $\Phi_H$s of the magnetic field and under different values of the electric field $E_b$. The field sweeps correspond to an adiabatic evolution of the system between locally stable or metastable minima of the free-energy landscape; no real-time dynamics is implied, and hysteresis effects arise from the presence of finite energy barriers separating these minima.
Fig.~\ref{fig:phaseDiag}{\bf A, B} shows representative free-energy landscapes at fixed $\Phi_H$, $E_b$ and $h$, with the white lines indicating the order parameter orientation trajectory $(\phi_1, \phi_2)$ during the sweeps. Fig.~\ref{fig:phaseDiag}{\bf C} was obtained by overlapping $P_b(E_b,\phi_H)$ plots obtained above and below the spin reorientation transition in subsequent sweeps $0T \leq h \leq 10T$. Fig.~\ref{fig:fig3}{\bf A-D} displays the evolution of $P_b\propto\cos(\phi_2 - \phi_1 )$ computed  during magnetic-field sweeps as discussed above, at different fixed $\Phi_H$ and $E_b$.

%
%
%
%
%
%




\newpage


\renewcommand{\thefigure}{S\arabic{figure}}
\renewcommand{\thetable}{S\arabic{table}}
\renewcommand{\theequation}{S\arabic{equation}}
\renewcommand{\thepage}{S\arabic{page}}
\setcounter{figure}{0}
\setcounter{table}{0}
\setcounter{equation}{0}
\setcounter{page}{1} 


\begin{center}
\section*{Supplementary Materials for\\ \scititle}

\author{
    M. Ryzhkov$^{1}$,
    A. Granero$^{6}$,
    J. Wettstein$^{1}$,
    Anna Pimenov$^{1}$,
    X.~Wang$^{2}$,
    L. Ponet$^{5}$,\and
    S.-W. Cheong$^{3}$,
    M. Mostovoy$^{4}$,
    Andrei Pimenov,
    S. Artyukhin$^{6}$,$^{1\ast}$\\
	\small$^{1}$Institute of Solid State Physics, Vienna University of Technology, 1040 Vienna, Austria.\and
    
    \small$^{2}$University of Science and Technology, Beijing, 
    China.\and
    
    \small$^{3}$Rutgers Center for Emergent Materials and Department of Physics and Astronomy, \and 
    
    \small Rutgers University, New Jersey 08854, USA.\and
    
    \small$^{4}$Zernike Institute for Advanced Materials, University of Groningen, the Netherlands.\and
    
    \small$^{5}$Theory and Simulation of Materials (THEOS) and National Centre for Computational Design\\and Discovery of Novel Materials (MARVEL),\and 
    
    \small École Polytechnique Fédérale de Lausanne, Lausanne 1015, Switzerland.\and
    
    \small$^{6}$Quantum Materials Theory, Istituto Italiano di Tecnologia, 16163 Genova, Italy.\and
    
	\small$^\ast$Corresponding author. Email: sergey.artyukhin@iit.it\and
}

\end{center}

\subsubsection*{This PDF file includes:}
{\bf Materials and Methods:}\\
Experimental\\
Landau free energy derivation\\
{\bf Supplementary figures:}\\
Figure S1: Estimating the spin reorientation field\\
Figure S2: Simulated switching between magnetoelectric states under $H$ sweeps and a constant $E$-field\\
{\bf Supplementary tables:}\\
Tables S1: Transformation properties of AFM order parameters \\

\newpage



\subsection*{Landau free energy and parameter estimation}
\begin{figure}[t]
\centering
\includegraphics[width=\linewidth]{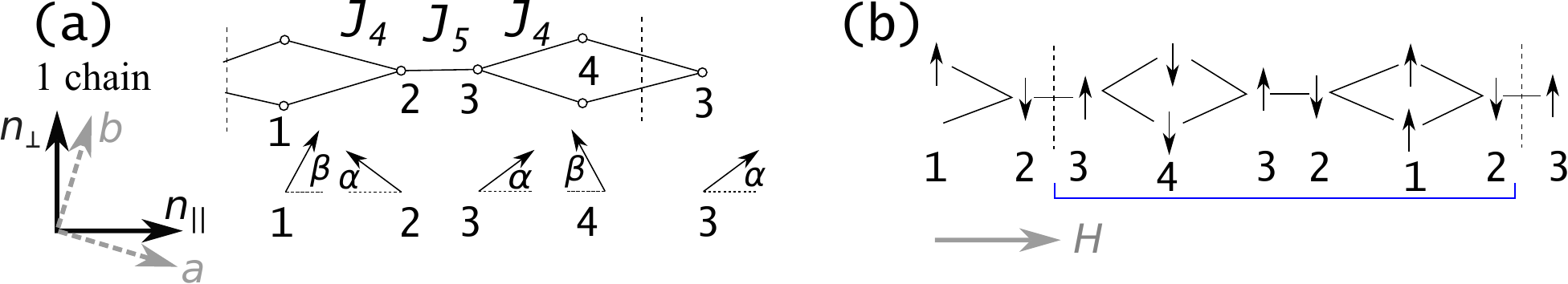}
    \caption{{\bf Estimating the spin reorientation field: } (a) Magnetic exchange constants and a spin configuration for a single chain. (b) A collinear spin configuration.}
    \label{fig:S1param}
\end{figure}

We use the transformation properties of the paramagnetic space group P$bam$ (\#55 in the International Tables~\cite{ITA2002}) summarized in Table ~\ref{tab:symmetry} to write the invariants that make up the Landau free energy:
\begin{equation}
        \mathcal{F}= \gamma(\bm{L}_1\cdot\bm{L}_2)^2-\alpha E_b(\bm{L}_1\cdot\bm{L}_2)+
    \sum_{i=1,2}\left[-K_2(\bm{L}_i\cdot \bm{n}_i)^2-K_4(\bm{L}_i\cdot \bm{n}_i')^4+(\bm{L}_i\cdot\bm{H})^2
    \right],
\end{equation}
$\bm{L}_1$ and $\bm{L}_2$ have opposite parity and therefore the invariant $\alpha E_b(\bm{L}_1\cdot\bm{L}_2)$ is allowed. The terms with coefficients $K_2$ and $K_4$ represent the most general single-ion anisotropy up to $\cos 4\phi_L$. In estimating the model parameters, we assume for simplicity that the easy axes of Mn$^{3+}$ and Mn$^{4+}$ are parallel. Exchange and anisotropy energy contributions for a single chain in the spin configuration, shown in Fig.~\ref{fig:S1param}, are:
\begin{multline}
    E_1=J_5 S_A^2(\bm{n}_2\cdot \bm n_3)+2J_4 S_A S_B(\bm n_1 \cdot \bm n_2 + \bm n_3\cdot \bm n_4)+K_AS_A^2(n_{2\perp}^2+n_{3\perp}^2)
    +K_BS_B^2(n_{1\perp}^2+n_{4\perp}^2)\\-2\mu_{\mathrm{B}}H_{\|}\left(S_A(n_{2\|}+n_{3\|})+S_B(n_{1\|}+n_{4\|})\right),
\end{multline}
where $S_A=2$ and $S_B=3/2$ stand for spin magnitudes at Mn$^{3+}$ and Mn$^{4+}$ sites, and $\bm n_i$ are unit vectors along the corresponding classical spins. Defining the orientations of $\bm n_{1,2}$ by angles $\alpha,\beta$, as shown in Fig.~\ref{fig:S1param}A, the energy takes the form:
\begin{equation}
    E_1=-J_5 S_A^2\cos 2\alpha-4J_4S_aS_B\cos(\alpha+\beta)+2K_AS_A^2\cos^2\alpha+2K_BS_B^2\cos^2\beta-4\mu_{\mathrm{B}}(S_A\sin\alpha+S_B\sin\beta).\label{eq:E1}
\end{equation}
Assuming that the SOC-induced anisotropy is much weaker than the exchange ($K\ll J$), we expand in small $\alpha, \beta$ to obtain 
\begin{equation}
    \alpha\approx \frac{\mu_{\mathrm{B}}H}{J_5}\frac{S_a-S_B}{S_A^2},\quad \beta=\frac{\mu_{\mathrm{B}}H}{J_4S_A}-\alpha\label{eq:ab}
\end{equation}
For a spin-flopped state at $H>H_{SF}$, shown in Fig.~\ref{fig:S1param}B, one finds $E=J_5S_A^2-4J_4S_AS_B$. Equating this energy at $H_{SF}$ to the energy before the spin flop, Eq.~\ref{eq:E1} with optimal $\alpha, \beta$, Eq.~\ref{eq:ab}, we find
\begin{equation}
    K_AS_A^2+K_BS_B^2=(\mu_{\mathrm{B}}H_{SF})^2
\left[\frac{1}{16J_5}+\frac{3}{4J_4}\right].
\end{equation}
Using the values from \cite{Vaunat2021}, $H_{SF}=5$~T, $J_4=2.25$~meV, $J_5=0.15$~meV, we obtain $K_AS_A^2+K_BS_B^2=0.73$~K. For $E=5\cdot 10^4$~V/cm, $P=3.5\cdot 10^3 \mu$C/m$^2$,  $EP=1.75\cdot 10^4$~J/m$^3$, $V_{uc}=360$~\AA, we find the magnetoelectric energy per chain in 1 unit cell $EPV/k_{\mathrm{B}}\approx0.1$~K. Anisotropy energy per chain in one unit cell is $2(K_AS_A^2+K_BS_B^2)=1.45$~K.

\begin{figure}
    \centering    \includegraphics[width=1\linewidth]{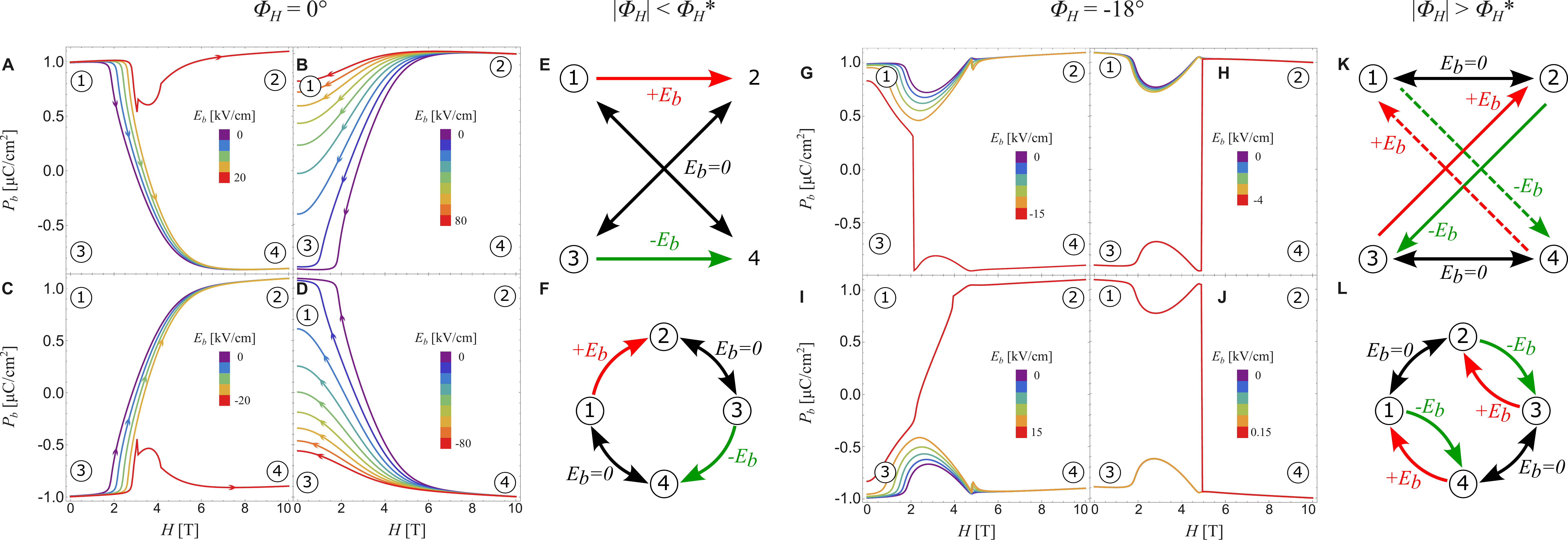}
    \caption{\label{fig:figS2}{\bf Simulated switching between magnetoelectric states under $H$ sweeps and a constant $E$-field.} Color encodes the magnitude of $E_b$. (\textbf{A,B}) Switching driven by sweeps of  magnetic field at $\Phi_H = 0^{\circ}$. (\textbf{A}) At low $E$-field switching follows $1\to 4$. $E$-field above the threshold results in $1\to 2$ instead. (\textbf{B}) Switching starting from state 2. Continuous deformation of $P(H)$ curves indicates that the final state is merely polarized by $E$-field. Only much larger fields $E_b\approx 100$~kV/cm change the final state to 1. (\textbf{C, D}) Switching paths starting in states 3 and 4 are obtained by a mirror $P\to -P, E\to -E, \bm L_1\to -\bm L_1$. (\textbf{G, H}) $E$-field controlled switching starting from states 1 and 2, respectively, when the $H$-field angle exceeds the topological switching angle interval, $\Phi_H = -18^{\circ}$. Here, a moderate $E$-field does change the switching path (red curves). Lower panels (\textbf{I, J}) illustrating switching from states 3,4, are obtained from the upper ones by mirror with respect to $P=0$. 
    (\textbf{E,F,K,L}) Switching paths, enabled by an external electric fields (red and green arrows) below (E,F) and above (K,L) the magic angle interval $7.7^{\circ} \leq |\Phi_H^*| \leq 17.7^{\circ}$.}    
\end{figure}

\begin{table}[b]
    \centering
    \begin{tabular}{c|c|c|c}
    &$m_z$ & I& $2_y$\\
    \hline
     $1$& $1$& $2-a$& $3$\\
     $2$& $2$& $1-a$& $4-a$\\ 
     $3$& $3$& $4-a$& $1$\\ 
     $4$& $4$& $3-a$& $2-a$\\ 
     $1,2$& $1,2$& $1,2$& $3,4$\\
     $3,4$& $3,4$& $3,4-a$& $1,2$\\ 
    $L_{1x}$& $-L_{1x}$& $-L_{1x}$& $-L_{2x}$\\ 
    $L_{1y}$& $-L_{1y}$& $-L_{1y}$& $L_{2y}$\\
    $L_{2x}$& $-L_{2x}$& $L_{2x}$& $-L_{1x}$\\
    $L_{2y}$& $-L_{2y}$& $L_{2y}$& $L_{1y}$\\
    $P_x$& $P_x$& $-P_x$& $-P_x$\\
    $P_y$& $P_y$& $-P_y$& $P_y$\\
    $P_z$& $-P_z$& $-P_z$& $-P_z$\\
    $\bm{L}_1\cdot \bm{L}_2$& $\bm{L}_1\cdot \bm{L}_2$& $-\bm{L}_1\cdot \bm{L}_2$& $\bm{L}_1\cdot \bm{L}_2$\\
     $[\bm{L}_1\times \bm{L}_2]_z$& $[\bm{L}_1\times \bm{L}_2]_z$& $-[\bm{L}_1\times \bm{L}_2]_z$& $-[\bm{L}_1\times \bm{L}_2]_z$\\
    $H_x$& $-H_x$& $H_x$& $-H_x$\\
    $H_y$& $-H_y$& $H_y$& $H_y$\\
    \end{tabular}
    \caption{\label{tab:symmetry}\textbf{Transformation properties of AFM order parameters} for zigzag Mn chains $\bm{L}_1, \bm{L}_2$ and of single ion spins (labeled according to Fig.~\ref{fig:fig1}) under the generators of the P$bam$ space group (\#55 in the International Tables of Crystallography).}
\end{table}
\begin{figure}
    \centering
\includegraphics[width=0.5\linewidth]{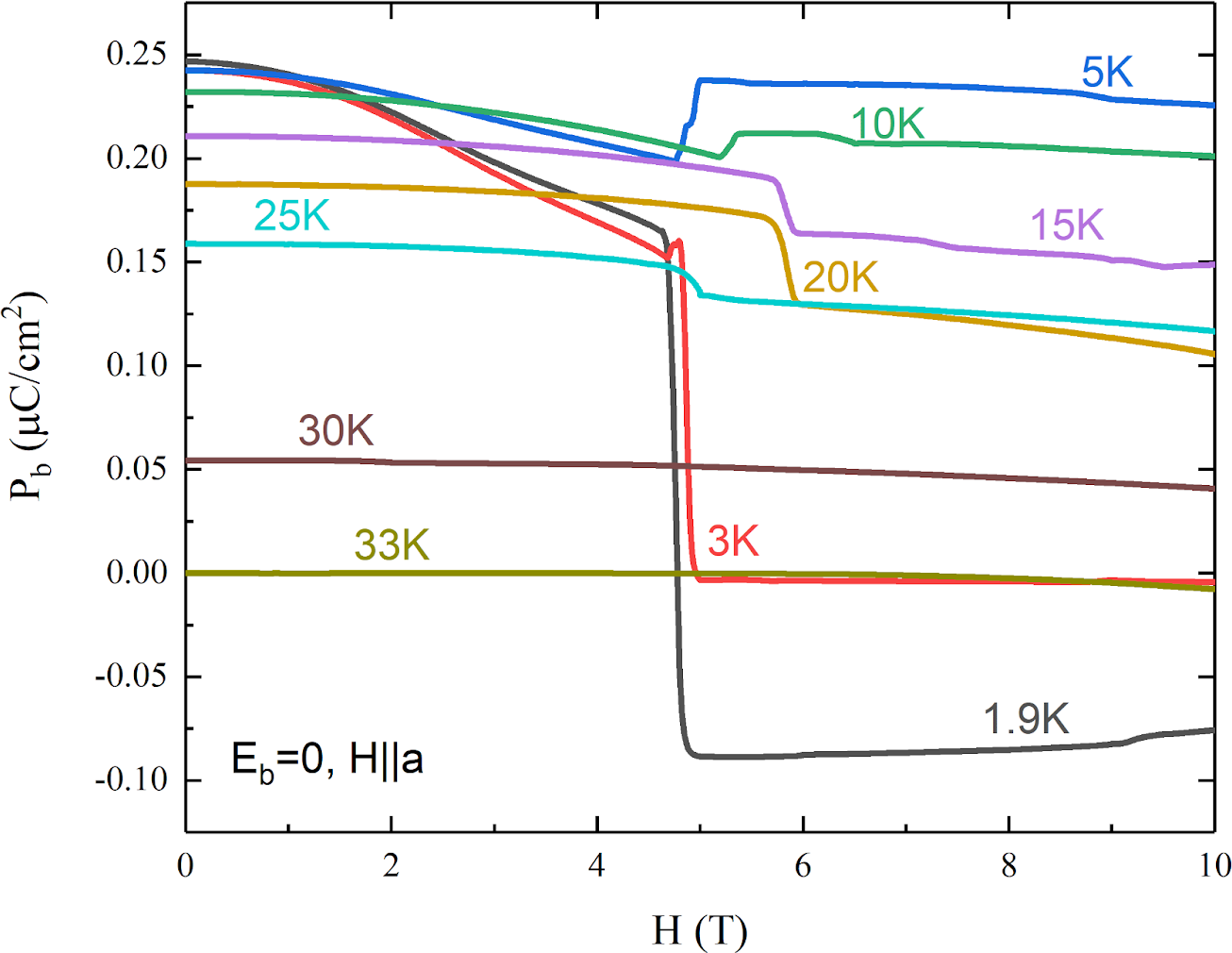}
    \caption{Temperature dependence of switching between states $1\to 4$ under $\bm{H}\| a$ and $E_b=0$. The switching persists up to 25~K.}
    \label{fig:SwitchT}
\end{figure}

\clearpage 

\clearpage
\end{document}